%% file: SMP-13-009_temp.tex
\pdfoutput=1

\documentclass[11pt,twoside,a4paper,cmspaper,final,collab]{cms-tdr}
\def\svnVersion{257862}\def\cmsCernNo{2013-037}\def\cmsCernDate{\today}\def\cmsMessage{Published in Physical Review D as \href{http://dx.doi.org/10.1103/PhysRevD.90.032008}{\doi{10.1103/PhysRevD.90.032008.}}}

\begin{document}\cmsNoteHeader{SMP-13-009}

\hyphenation{had-ron-i-za-tion}
\hyphenation{cal-or-i-me-ter}
\hyphenation{de-vices}

\RCS$Revision: 249606 $
\RCS$HeadURL: svn+ssh://svn.cern.ch/reps/tdr2/papers/SMP-13-009/trunk/SMP-13-009.tex $
\RCS$Id: SMP-13-009.tex 249606 2014-07-03 16:22:19Z alverson $
\newlength\cmsFigWidth
\ifthenelse{\boolean{cms@external}}{\setlength\cmsFigWidth{0.95\columnwidth}}{\setlength\cmsFigWidth{0.7\textwidth}}
\ifthenelse{\boolean{cms@external}}{\providecommand{\cmsLeft}{top}}{\providecommand{\cmsLeft}{left}}
\ifthenelse{\boolean{cms@external}}{\providecommand{\cmsRight}{bottom}}{\providecommand{\cmsRight}{right}}
\ifthenelse{\boolean{cms@external}}{\providecommand{\NA}{\ensuremath{\cdots}}}{\providecommand{\NA}{\text{---}}}
\ifthenelse{\boolean{cms@external}}{\providecommand{\CL}{C.L.\xspace}}{\providecommand{\CL}{CL\xspace}}
\newcolumntype{C}{D{,}{{}\pm{}}{-1}}
\cmsNoteHeader{SMP-13-009} 
\newcommand{\mjj}{\ensuremath{m_{jj}}\xspace}%
\title{\texorpdfstring{A search for $\PW\PW\gamma$ and $\PW\Z\gamma$ production and constraints
on anomalous quartic gauge couplings in $\Pp\Pp$ collisions at $\sqrt{s}=8$\TeV}{A search for WW gamma and WZ gamma production and constraints on anomalous quartic gauge couplings in pp collisions at sqrt(s) = 8 TeV}}

\date{\today}

\abstract{
   A search for ${\PW}V\gamma$ triple vector boson production is presented
   based on events containing a {\PW} boson decaying to a muon or an electron and a neutrino,
   a second $V$ ({\PW} or {\Z}) boson, and a photon.
   The data correspond to an integrated luminosity of 19.3\fbinv
   collected in 2012 with the CMS detector at the LHC in $\Pp\Pp$ collisions at $\sqrt{s}=8\TeV$.
   An upper limit of 311\unit{fb} on the cross section for the ${\PW}V\gamma$ production
   process is obtained at 95\%
   confidence level for photons with a transverse energy above 30\GeV and with an absolute value
   of pseudorapidity of less than 1.44. This limit is approximately a factor of 3.4 larger than
   the standard model predictions that are based on next-to-leading order
   QCD calculations. Since no evidence of anomalous $\PW\PW\gamma\gamma$ or $\PW\PW\Z\gamma$
   quartic gauge boson couplings is found, this paper presents the first experimental limits on
   the dimension-8 parameter
   $f_{T,0}$ and the CP-conserving $\PW\PW\Z\gamma$ parameters $\kappa_0^\PW$ and $\kappa_C^\PW$. Limits
   are also obtained for the $\PW\PW\gamma\gamma$ parameters $a_{0}^\PW$ and $a_{C}^\PW$.
}

\hypersetup{%
pdfauthor={CMS Collaboration},%
pdftitle={A search for WW gamma and WZ gamma production and constraints on anomalous quartic gauge couplings in pp collisions at sqrt(s) = 8 TeV},%
pdfsubject={CMS},%
pdfkeywords={CMS, physics, gauge couplings}}

\maketitle 
\clearpage{}
\section{Introduction}
\label{sec:intro}

The standard model (SM) of particle physics provides a good description of the existing
high-energy data~\cite{Beringer:1900zz}. The diboson {\PW\PW} and {\PW\Z} production
cross sections have been precisely measured at the Large Hadron Collider (LHC) and are in agreement
with SM expectations~\cite{CMSdiboson,Chatrchyan:2012bd,Chatrchyan:2013yaa,ATLAS:2012mec,Aad:2012twa}.
This paper presents a search for the production of three gauge bosons $\PW\PW\gamma$
and $\PW\Z\gamma$, together denoted as ${\PW}V\gamma$. It represents an extension of the
measurement of diboson production presented in Ref.~\cite{Chatrchyan:2012bd}, with the
additional requirement of an energetic photon in the final state. Previous searches for
triple vector boson production, when at least two bosons are massive, were performed at
LEP~\cite{wwaLEP:1999,Achard:2001eg,Abdallah:2003xn,Heister:2004yd,Schael:2013ita}.

The structure of gauge boson self-interactions emerges naturally in the SM from the
non-Abelian $SU(2)_{L} \otimes U(1)_{Y}$ gauge symmetry. Together with the triple
${\PW}V\gamma$ gauge boson vertices, the SM also predicts the existence of the quartic
$\PW\PW\PW\PW$, $\PW\PW\Z\Z$, $\PW\PW\Z\gamma$, and $\PW\PW\gamma\gamma$ vertices.  The
direct investigation of gauge boson self-interactions provides a crucial test of the
gauge structure of the SM, and one that is all the more significant at LHC
energies~\cite{Belyaev:1998ih}.

The study of gauge boson self-interactions may also provide evidence for the
existence of new phenomena at a higher energy
scale~\cite{aihara1996,Du:2012vh,Fichet:2013ola,Giudice:2007fh}. Possible new physics
beyond the SM, expressed in a model independent way by higher-dimensional effective
operators~\cite{500GeVNLC,Belanger:1999, Bosonic:2004PRD, Eboli:2006wa,Yang:2012vv,Ye:2013psa},
can be implemented with anomalous triple gauge and quartic gauge couplings (AQGC), both
of which contribute in triple gauge boson production. A deviation of one of the
couplings from the SM prediction could manifest itself in an enhanced production
cross section, as well as a change in the shape of the kinematic distributions of the
${\PW}V\gamma$ system. CMS recently obtained a stringent limit on the anomalous
$\PW\PW\gamma\gamma$ quartic coupling via the exclusive two-photon production of
{\PWp\PWm}~\cite{Chatrchyan:2013foa}.

This paper presents a search for ${\PW}V\gamma$ production in the single lepton final
state, which includes $\PW(\to\ell\Pgn)\PW(\to jj)\gamma$ and
$\PW(\to\ell\Pgn)\Z(\to jj)\gamma$ processes, with $\ell = \Pe,\Pgm$. The data used
in this analysis correspond to a total integrated luminosity of
$19.3 \pm 0.5$ $(19.2 \pm 0.5)\fbinv$~\cite{lumiPAS} collected with the CMS detector
in the muon (electron) channel in $\Pp\Pp$ collisions at $\sqrt{s} = 8\TeV$ in 2012. The
hadronic decay mode is chosen because the branching fraction is substantially higher
than that of the leptonic mode. However, the two production processes $\PW\PW\gamma$
and $\PW\Z\gamma$ cannot be clearly differentiated since the detector dijet mass resolution
($\sigma\sim 10$\%)~\cite{Chatrchyan:2011ds} is comparable to the mass difference between
the {\PW} and {\Z} bosons. Therefore, $\PW\PW\gamma$ and $\PW\Z\gamma$ processes are
treated as a single combined signal.

\section{Theoretical framework}
\label{sec:aQGCth}

An effective field theory approach is adopted in which
higher-dimensional operators supplement the SM Lagrangian to include
anomalous gauge couplings. Within this
framework, anomalous boson interactions can be parametrized using two
possible representations. The first is a nonlinear realization of the
$SU(2)\otimes U(1)$ gauge symmetry that is broken by means other
than the conventional Higgs scalar doublet~\cite{Belanger:1999,Bosonic:2004PRD}. The quartic boson
interactions involving photons appear as dimension-6 operators. The
second is a linear realization of the symmetry that is broken by the
conventional Higgs scalar doublet~\cite{Belanger:1999,Eboli:2006wa}. The quartic interactions
involving photons appear as dimension-8 operators.

Some of the operators within one realization share similar Lorentz
structures with operators from the other,
so that their parameters can be expressed simply in terms of each
other, whereas others cannot.
While the discovery of the SM Higgs boson makes the linear realization
more appropriate for AQGC searches~\cite{aihara1996,Eboli:2006wa},
it contains 14 such operators that can contribute to the
anomalous coupling signal. In addition, all published AQGC limits to
date are expressed in terms of dimension-6 parameters. To bridge this
divide, we select four dimension-6 parameters, two of which have not
been previously measured, and the other two are used to compare with
previous results~\cite{Belanger:1999,Achard:2001eg}. These parameters
also have dimension-8 analogues. Finally, we include a representative
parameter from the linear realization, $f_{T,0}$, which has no
dimension-6 analogue.

\begin{figure}[htb]
  \centering
      \includegraphics[width=0.45\columnwidth]{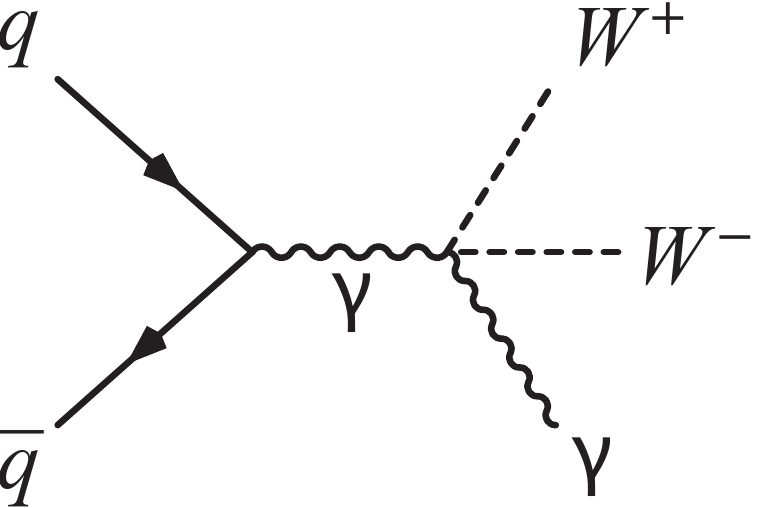}
      \includegraphics[width=0.45\columnwidth]{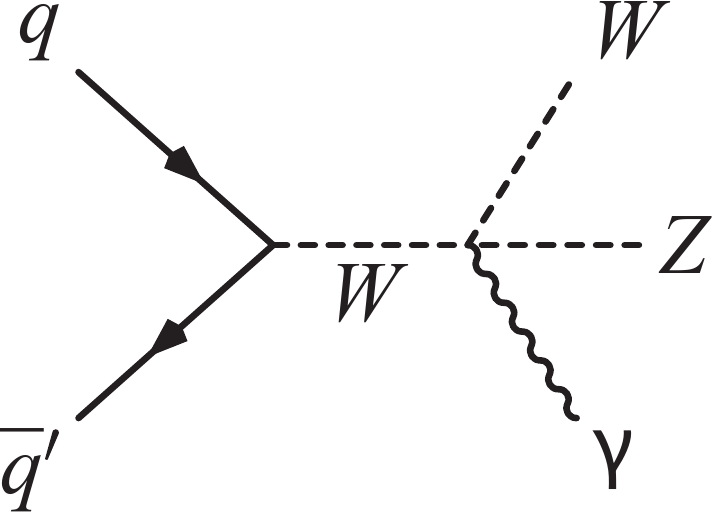}
    \caption{Feynman diagrams that involve a quartic vector boson vertex. Both diagrams are present in the SM.}
    \label{fig:feyndiag}
\end{figure}

The Feynman diagrams for the quartic vertices are shown in
Figure~\ref{fig:feyndiag}, and the CP-conserving, anomalous
interaction Lagrangian terms chosen for this analysis are written in
Eq.~\eqref{lagrangian}.

\ifthenelse{\boolean{cms@external}}{
\begin{multline}\label{lagrangian}
\mathcal{L}_{\textrm{AQGC}} = -\frac{e^2}{8} \frac{a_0^\PW}{\Lambda^2} F_{\mu \nu} F^{\mu \nu} W^{+\alpha} W^-_{\alpha} \\- \frac{e^2}{16}
\frac{a_C^\PW}{\Lambda^2} F_{\mu \nu} F^{\mu \alpha} ( W^{+\nu} W^-_{\alpha} + W^{-\nu} W^+_{\alpha} ) \\
 - e^2 g^2 \frac{\kappa_0^\PW}{\Lambda^2} F_{\mu \nu} Z^{\mu \nu} W^{+ \alpha} W^-_{\alpha}\\ - \frac{e^2 g^2}{2} \frac{\kappa_C^\PW}{\Lambda^2} F_{\mu \nu} Z^{\mu \alpha} ( W^{+\nu} W^-_{\alpha} + W^{-\nu} W^+_{\alpha} )\\
 + \frac{f_{T,0}}{\Lambda^4} \Tr[\hat{W}_{\mu \nu} \hat{W}^{\mu \nu}] \times \Tr[\hat{W}_{\alpha \beta} \hat{W}^{\alpha \beta}].
\end{multline}
}{
\begin{equation}\begin{aligned}\label{lagrangian}
\mathcal{L}_{\textrm{AQGC}} =& -\frac{e^2}{8} \frac{a_0^\PW}{\Lambda^2} F_{\mu \nu} F^{\mu \nu} W^{+\alpha} W^-_{\alpha} - \frac{e^2}{16}
\frac{a_C^\PW}{\Lambda^2} F_{\mu \nu} F^{\mu \alpha} ( W^{+\nu} W^-_{\alpha} + W^{-\nu} W^+_{\alpha} ) \\
 &- e^2 g^2 \frac{\kappa_0^\PW}{\Lambda^2} F_{\mu \nu} Z^{\mu \nu} W^{+ \alpha} W^-_{\alpha}- \frac{e^2 g^2}{2} \frac{\kappa_C^\PW}{\Lambda^2} F_{\mu \nu} Z^{\mu \alpha} ( W^{+\nu} W^-_{\alpha} + W^{-\nu} W^+_{\alpha} ) \\
 &+ \frac{f_{T,0}}{\Lambda^4} \Tr[\hat{W}_{\mu \nu} \hat{W}^{\mu \nu}] \times \Tr[\hat{W}_{\alpha \beta} \hat{W}^{\alpha \beta}].
\end{aligned}\end{equation}
}
The energy scale of possible new physics is represented by $\Lambda$,
$g = e/\sin(\theta_\PW)$, $\theta_\PW$ is the Weinberg angle, $e$ is the
unit of electric charge, and the usual field tensors are defined in
Ref.~\cite{Eboli:2006wa,Belanger:1999,Bosonic:2004PRD}.
The dimension-6 parameters $a_0^\PW/\Lambda^2$ and $a_C^\PW/\Lambda^2$ are
associated with the $\PW\PW\gamma\gamma$ vertex and the
$\kappa_0^\PW/\Lambda^2$ and $\kappa_C^\PW/\Lambda^2$ parameters are
associated with the $\PW\PW\Z\gamma$ vertex. The dimension-8 parameter
$f_{T,0}/\Lambda^4$ contributes to both vertices. The
$a_{0,C}^\PW/\Lambda^2$ coupling parameters have
dimension-8 analogues, the $f_{M,i}/\Lambda^4$ coupling parameters. The
relationship between the two is as follows~\cite{Belanger:1999}(Eq.~3.35):
\begin{equation}\begin{aligned}
\frac{a_0^\PW}{\Lambda^2} &= -\frac{4 M_\PW^2}{g^2} \frac{f_{M,0}}{\Lambda^{ 4}} - \frac{8 M_\PW^2}{{g'}^2} \frac{f_{M,2}}{\Lambda^{ 4}},\\
\frac{a_C^\PW}{\Lambda^2} &= \frac{4 M_\PW^2}{g^2} \frac{f_{M,1}}{\Lambda^{ 4}} + \frac{8 M_\PW^2}{{g'}^2} \frac{f_{M,3}}{\Lambda^{ 4}},
\label{dim6to8}
\end{aligned}
\end{equation}
where $g' = e/\cos(\theta_\PW)$ and $M_\PW$ is the invariant mass of the $\PW$ boson.
The expressions listed in Eq.~\eqref{dim6to8} are used to translate
the AQGC limits obtained for $a_{0,C}^\PW/\Lambda^2$, into limits on $f_{M,i}/\Lambda^4$.
It is also required that $f_{M,0} = 2 \times f_{M,2}$ and $f_{M,1} = 2 \times f_{M,3}$, which
results in the suppression of the contributions to the $\PW\PW\Z\gamma$ vertex in
Eq.~\eqref{dim6to8}, as can be seen from~\cite{Bosonic:2004PRD} Eq.~22 and Eq.~23.

Any nonzero value of the AQGCs will lead to tree-level unitarity
violation at sufficiently high energy. We find that the unitarity
condition~\cite{AQGC:2001} cannot be generally satisfied by the addition of a dipole
form factor; however, unitarity conserving new physics with a
structure more complex than that represented by a dipole form
factor is possible. Since the structure of new physics is
not known \textit{a priori}, the choice is made to set limits without
using a form factor.

\section{The CMS detector }
\label{sec:CMSdet}

The central feature of the Compact Muon Solenoid (CMS) apparatus is a
superconducting solenoid of 6\unit{m} internal diameter and 13\unit{m} length, providing a
magnetic field of 3.8\unit{T}. Within the superconducting solenoid
volume are a silicon pixel and strip tracker, a lead tungstate crystal
electromagnetic calorimeter (ECAL), and a brass/scintillator hadron
calorimeter (HCAL). Muons are reconstructed in gas-ionization detectors
embedded in the steel flux-return yoke outside the solenoid. Extensive
forward calorimetry complements the coverage provided by the brass/scintillator section of the hadronic calorimeter.

The CMS experiment uses a right-handed coordinate system, with the origin at the
nominal interaction point, the $x$ axis pointing to the center of the
LHC ring, the $y$ axis pointing up (perpendicular to the LHC plane), and
the $z$ axis along the counterclockwise beam direction. The polar angle
$\theta$ is measured from the positive $z$ axis and the azimuthal
angle $\phi$ is measured in radians in the $x$-$y$ plane.
The pseudorapidity $\eta$ is defined as $\eta = -\ln[\tan(\theta/2)]$.

The energy resolution for photons with transverse energy (\ET) of 60\GeV varies between 1.1\% and 2.6\% in the ECAL
barrel, and from 2.2\% to 5\% in the endcaps~\cite{Chatrchyan:2013dga}. The HCAL, when combined with the ECAL, measures jets
with a resolution $\Delta E/E \approx 100\% / \sqrt{E\,[\GeVns]} \oplus 5\%$ \cite{CMS:2011esa}. To improve
the reconstruction of jets, the tracking and calorimeter information is
combined using a particle flow (PF) reconstruction technique~\cite{PFT-09-001}. The jet energy resolution
typically amounts to 15\% at 10\GeV, 8\% at 100\GeV, and 4\% at 1\TeV.

A more detailed description of the CMS detector can be found in Ref.~\cite{Chatrchyan:2008zzk}.

\section{Event simulation }
\label{sec:mc}
All Monte Carlo (MC) simulation samples, except for the single-top-quark samples, are generated with the
 {\MADGRAPH} 5.1.3.22~\cite{MadGraph} event generator using the CTEQ6L1 parton
distribution functions (PDF). Single-top-quark samples are generated with {\POWHEG} (v1.0, r1380)~\cite{Nason:2004rx,Frixione:2007vw,Alioli:2010xd,Alioli:2009je,Re:2010bp} with the CTEQ6M
PDF set~\cite{Pumplin:2002vw,Nadolsky:2008zw}.
The matrix element calculation is used, and outgoing partons are matched to parton showers from
{\PYTHIA} 6.426~\cite{Sjostrand:2006za} tune $Z2^*$~\cite{Collaboration:2012tb} with
a matching threshold of 20\GeV and a dynamic factorization ($\mu_F$) and
renormalization ($\mu_R$) scale given by $\sqrt{\smash[b]{m^{2}_{W/Z} + p^{2}_{T,W/Z}}}$.
The next-to-leading-order/leading-order (NLO/LO) QCD cross section correction
factors (K-factors) for ${\PW}V\gamma$ and AQGC diagrams
are derived using the NLO cross sections calculated with {a\MCATNLO}~\cite{Frixione:2002ik}. The MSTW2008nlo68cl~\cite{Martin:2009iq} PDF
set is used to calculate the PDF uncertainty following the prescription of Ref.~\cite{Frederix:2011ss}.
The K-factor obtained for ${\PW}V\gamma$ is consistent with a
constant value of 2.1 for photons with $\ET > 30\GeV$ and $\abs{\eta^{\gamma}} < 2.5$. The K-factor for AQGC diagrams is
found to be close to 1.2.
A summary of the contributing processes and their cross section is given in Table~\ref{tab:samples}.

\begin{table}[htb]
  \topcaption{\label{tab:samples}
Cross sections used to normalize the simulated samples. All cross sections are given for a photon $\ET > 10\GeV$, $|\eta^{\gamma}| < 2.5$.
The order of the cross section calculation is also indicated. The normalization for the $\PW\gamma$+jets sample is derived from data. }
\centering
  \begin{scotch}{lrC}

  Process & &\multicolumn{1}{l}{Cross section [pb]} \\
  \hline
  SM $\PW\PW\gamma$                    & (NLO) &    0.090, 0.021    \\
  SM $\PW\Z\gamma$                     & (NLO) &    0.012, 0.003    \\
  \hline
  $\PW\gamma$ + jets                   & (Data)&    10.9, 0.8    \\
  $\Z\gamma$ + jets                    & (LO)  &    0.63, 0.13 \\
  $\ttbar\gamma$                       & (LO)  &    0.62, 0.12 \\
  Single $\cPqt$ + $\gamma$(inclusive) & (NLO) &    0.31, 0.01  \\
  \end{scotch}
\end{table}

To simulate the signal events for a given AQGC parameter set, several samples are generated with a range of parameter values
and the other AQGC parameters are set to zero.

A \GEANTfour-based simulation~\cite{GEANT4} of the CMS
detector is used in the production of all MC samples. All simulated events are reconstructed and analyzed
with the same algorithms that are used for the LHC collision events.
Additional corrections (scale factors) are applied to take into account the difference in lepton
reconstruction and identification efficiencies observed between data and simulated events.
For all simulated samples, the hard-interaction collision is overlaid with the appropriate number of simulated
minimum bias collisions. The resulting events are weighted to reproduce the distribution of
the number of inelastic collisions per bunch crossing (pileup) inferred from data.

\section{Event reconstruction and selection}
\label{sec:recosel}

The data used in this analysis corresponds to a total integrated
luminosity of $19.3 \pm 0.5$ $(19.2 \pm 0.5)\fbinv$~\cite{lumiPAS} collected with
the CMS detector in the muon (electron) channel in $\Pp\Pp$ collisions at
$\sqrt{s} = 8\TeV$ in 2012.  The data were recorded with single-lepton
triggers using \pt thresholds of 24\GeV for muons and 27\GeV for
electrons. The overall trigger efficiency is about 94\% (90\%) for
muon (electron) data, with a small dependence (a few percent) on \pt
and $\eta$. Simulated events are corrected for the trigger efficiency
as a function of lepton \pt and $\eta$.

The events used in this analysis are characterized by the production
of a photon plus a pair of massive gauge bosons ({\PW\PW} or {\PW\Z}),
where one {\PW} boson decays to leptons and the other boson ({\PW} or
{\Z}) decays to quarks. To select leptonic {\PW} boson decays, we
require either one muon ($\pt > 25\GeV$, $\abs{\eta} < 2.1$) or one
electron ($\pt > 30\GeV$, $\abs{\eta} < 2.5$, excluding the transition
region between the ECAL barrel and endcaps $1.44 < \abs{\eta} < 1.57$ because
the reconstruction of an electron in this region is not optimal). 
The offline lepton $\pt$ thresholds is set in the stable, high-efficiency
region above the corresponding trigger thresholds.
Events with additional leptons with $\pt > 10\,(20)\GeV$ for muons(electrons) are vetoed in order to
reduce backgrounds. The escaping neutrino results in missing
transverse energy (\ETslash) in the reconstructed event. Therefore a
selection requirement of $\ETslash > 35\GeV$ is applied to reject the
multijet backgrounds. The reconstructed transverse mass of the
leptonically decaying {\PW}, defined as
$\sqrt{\smash[b]{{\pt^{\ell}\ETslash[1-\cos(\Delta\phi_{\ell, \ETslash})]}}}$, where
$\Delta\phi_{\ell, \ETslash}$ is the azimuthal angle between the lepton and
the $\ETslash$ directions, is then required to exceed
30\GeV~\cite{WZCMS:2010}.  At least two jet candidates are required to
satisfy $\pt > 30\GeV$ and $\abs{\eta} < 2.4$.  The highest $\pt$ jet
candidates are chosen to form the hadronically decaying boson with mass
$m_{jj}$.  The photon candidate must satisfy $\ET > 30\GeV$ and
$\abs{\eta} < 1.44$.  Events with the photon candidate in one of the
endcaps ($\abs{\eta} > 1.57$) are excluded from the selection because
their signal purity is lower and systematic uncertainties are larger.

Jets and \ETslash~\cite{Chatrchyan:2011tn,WZCMS:2010} are formed from particles reconstructed using the PF algorithm.
Jets are formed with the anti-\kt clustering
algorithm~\cite{Cacciari:2008gp} with a distance parameter of 0.5.
Charged particles with tracks not originating from
the primary vertex are not considered for jet
clustering~\cite{Cacciari:2007fd, Cacciari:2008gn}. The primary vertex of the event is chosen to be the vertex with the highest
$\sum \pt^2$ of its associated tracks.
Jets are required to satisfy
identification criteria that eliminate candidates originating from noisy
channels in the hadron calorimeter~\cite{Chatrchyan:2009hy}.  Jet
energy corrections~\cite{Chatrchyan:2011ds} are applied to
account for the jet energy response as a function of $\eta$ and \pt,
and to correct for contributions from event pileup. Jets from pileup are
identified and removed using the trajectories of tracks associated with the jets,
the topology of the jet shape and the constituent multiplicities~\cite{Cacciari:2007fd,Cacciari:2008gn}.

The azimuthal separation between the highest \pt jet and the \ETslash
direction is required to be larger
than 0.4 radians. This criterion reduces the QCD multijet background where
the \ETslash can arise from a mismeasurement of the leading jet energy. To
reduce the background from $\PW\gamma$+jets events, requirements on the
dijet invariant mass $70 < m_{jj} < 100\GeV$, and on the
separation between the jets of $\abs{\Delta\eta_{jj}} < 1.4$, are
imposed. In order to reject top-quark backgrounds, the two jets are
also required to fail a {\cPqb} quark jet tagging requirement. The combined
secondary vertex algorithm~\cite{Chatrchyan:2012jua} is used, with a
discriminator based on the displaced vertex expected from {\cPqb} hadron
decays. This algorithm selects {\cPqb} hadrons with about 70\% efficiency,
and has a 1\% misidentification probability. The anti-\cPqb\ tag requirement suppresses approximately 7\% of the $\PW\PW\gamma$
and 10\% of the $\PW\Z\gamma$ signal via the $\PW\to \cPqc\cPaqs$,
$\cPZ\to\cPqb\cPaqb$ and $\cPZ\to\cPqc\cPaqc$ decays. These effects are taken into account in the analysis.

Muon candidates are reconstructed by combining information from the
silicon tracker and from the muon detector by means of a global track
fit. The muon candidates are required to pass the standard CMS muon
identification and the track quality criteria~\cite{muReco}. The isolation
variables used in the muon selection are based on the PF
algorithm and are corrected for the contribution from pileup. The muon
candidates have a selection efficiency of approximately 96\%.

Electrons are reconstructed from clusters~\cite{Chatrchyan:2013dga,Baffioni:2006cd,CMS:2011hqb,CMS:2013hoa} of ECAL energy
deposits matched to tracks in the silicon tracker within the ECAL
fiducial volume, with the exclusion of the transition region between the barrel and
the endcaps previously defined. The electron candidates are
required to be consistent with a particle originating from the primary
vertex in the event. The isolation variables used in the electron selection are based on the PF
algorithm and are corrected for the contribution from pileup.
The electron selection efficiency is approximately
80\%. To suppress the $\Z\to \Pep\Pem$ background in the electron
channel, where one electron is misidentified as a photon, a {\Z} boson mass veto of $\abs{M_{\Z} - m_{\Pe\gamma}} >
10\GeV$ is applied. The impact on the signal efficiency from applying such a suppression is negligible.

Photon candidates are reconstructed from clusters of cells with
significant energy deposition in the ECAL.
The candidates are required to be within the ECAL barrel fiducial region ($\abs{\eta} < 1.44$).
The observables used in the photon selection are isolation variables based on the PF algorithm
and they are corrected for the contribution due to pileup, the ratio of hadronic energy in the HCAL
that is matched in $(\eta,\phi)$ to the electromagnetic energy in the ECAL, the transverse
width of the electromagnetic shower, and an electron track veto.

\section {Background modeling}
\label{sec:BackgroundModeling}

The main background contribution arises from $\PW\gamma$+jets
production. After imposing the requirements described above, a binned
maximum likelihood fit to the dijet invariant mass distribution
\mjj~of the two leading jets is performed.  The signal region
corresponding to the {\PW} and {\Z} mass windows, $70 < \mjj< 100\GeV$,
is excluded from the fit. The contamination from ${\PW}V\gamma$ processes outside of the signal 
region is less than 1\%. The shape of the $\PW\gamma$+jets \mjj distribution is obtained from
simulation, and the normalization of this background component is unconstrained in the fit.
The normalization of the contribution from misidentified photons is allowed to float within a
Gaussian constraint of 14\% (Section~\ref{sec:Syst}). The post-fit ratio $K =
\sigma_{\text{fit}}/\sigma_{\mathrm{LO}}$ for the W$\gamma$+jets
background is $1.10 \pm 0.07$ ($1.07 \pm 0.09$) in the muon (electron) channel.

The background from misidentified photons arises mainly from the
\PW+3~jets process, where one jet passes the photon identification
criteria. The total contribution from misidentified photons is
estimated using a data control sample, where all selection criteria
except for the isolation requirement are applied. The shower shape
distribution is then used to estimate the total rate of misidentified
photons. Details on the method can be found in Ref.~\cite{Vgamma}. The
fraction of the total background from misidentified photons decreases
with photon \ET from a maximum of 23\% ($\pt = 30\GeV$) to 8\%
($\pt > 135\GeV$).

The multijet background is due to misidentified leptons from jets
that satisfy the muon or electron selection requirements. It is estimated by using a two component
fit to the \ETslash distribution in data. The procedure is described in~\cite{Chatrchyan:2012bd}, and was
repeated for the 8\TeV data. The multijet contribution is estimated
to be 6.2\% for the electron channel, with a 50\% uncertainty, and is negligible for the muon channel.

Other background contributions arise from top-quark pair production,
single-top-quark production, and $\Z\gamma$+jets. These are taken from
simulation and are fixed to their SM expectations, with the central
values and uncertainties listed in Table~\ref{tab:samples}. The
top-quark pair process contribution comes from the presence of two {\PW}
bosons in the decays. The
$\Z\gamma$+jets background can mimic the signal when the {\Z} decays
leptonically and one of the leptons is lost, resulting in
\ETslash. The sum of the top-quark pair, single-top-quark, and
$\Z\gamma$+jets backgrounds represent about 8\% of the expected SM
background rate.

\section{Systematic uncertainties}
\label{sec:Syst}

The uncertainties contributing to the measured rate of misidentified
photons arise from two sources. First, the statistical uncertainty is
taken from pseudo experiments drawn from the data control sample
described in Section~\ref{sec:BackgroundModeling} and is estimated to
be 5.6\% rising to 37\% with increasing photon \ET. The second arises
from a bias in the shower shape of {\PW}+3~jets simulation due
to the inverted isolation requirements. This uncertainty is estimated
to be less than 11\%. The combined uncertainty on the photon
misidentification rate, integrated over the \ET spectrum, is 14\%.

The uncertainty in the measured value of the
luminosity~\cite{lumiPAS} is 2.6\% and it contributes to the signal and
those backgrounds that are taken from the MC prediction. Jet energy
scale uncertainties contribute via selection thresholds on the
jet \pt and dijet invariant mass by 4.3\%. The small difference in \ETslash
resolution~\cite{Chatrchyan:2011tn} between data and simulation
affects the signal selection efficiency by less than 1\%. Systematic
uncertainties due to the trigger efficiency in the data (1\%) and
lepton reconstruction and selection efficiencies (2\%) are also accounted
for. Photon reconstruction efficiency and energy scale uncertainties contribute to the signal selection efficiency at the 1\% level. The uncertainty
from the {\cPqb} jet tagging procedure is 2\% on the data/simulation
efficiency correction factor~\cite{Chatrchyan:2012jua}. This
has an effect of 11\% on the $\ttbar\gamma$ background, 5\% on the single-top-quark
background, and a negligible effect on the signal. The theoretical
uncertainty in the $\ttbar\gamma$ and
$\Z\gamma$+jets production is 20\%.

The theoretical uncertainties in the
$\PW\PW\gamma$, $\PW\Z\gamma$, and AQGC signal cross sections are evaluated
using {\sc aMC@NLO} samples. We vary the renormalization and factorization
scales each by factors of 1/2 and 2, and require $\mu_R = \mu_F$, as described in
Ref.~\cite{Frederix:2011ss}. We find that the scale-related uncertainties are
23\%, and that the uncertainty due to the choice of PDF is 3.6\%.

\section{Upper limit on the standard model \texorpdfstring{${\PW}V\gamma$}{WV gamma} cross section}
\label{sec:xsec}

The SM ${\PW}V\gamma$ search is formulated as a simple counting experiment. The selected numbers of candidate events
in the data are 183 (139) in the muon (electron) channel. The predicted
number of background plus signal events is $194.2 \pm 11.5$ ($147.9 \pm 10.7$) in the muon (electron) channel, where the
uncertainty includes statistical, systematic and luminosity related uncertainties.
The event yield per process is summarized in Table~\ref{tab:evt}.

\begin{table}[htb]
\centering
  \topcaption{Expected number of events for each process. The predicted number of events for the $\PW\gamma$+jets and ${\PW}V$+jet processes, where the
jet is reconstructed as a photon, are derived from data. The "Total prediction" item represents the sum of all the individual contributions.}
  \begin{scotch}{lCC}
  \newlength\noe\settowidth{\noe}{number of events}
  Process 
  & \multicolumn{1}{l}{Muon channel} & \multicolumn{1}{l}{Electron channel}\\
  & \multicolumn{1}{l}{number of events} & \multicolumn{1}{l}{number of events}\\
  \hline
  SM $\PW\PW\gamma$                       & 6.6,1.5  & 5.0,1.1  \\
  SM $\PW\Z\gamma$                       & 0.6,0.1  & 0.5,0.1  \\
  \hline
  $\PW\gamma$ + jets                    & 136.9,10.5 & 101.6,8.5  \\
  ${\PW}V$ + jet, jet $\to \gamma$      & 33.1,4.8  & 21.3,3.3  \\
  MC $\ttbar\gamma$                     & 12.5,3.0  & 9.1,2.2  \\
  MC single top quark                   & 2.8,0.8  & 1.7,0.6  \\
  MC $\Z\gamma$ + jets                 & 1.7,0.1  & 1.5,0.1  \\
  Multijets                             &  \multicolumn{1}{c}{\NA}  & 7.2,5.1  \\
  \hline
    Total prediction                   & 194.2,11.5  & 147.9,10.7
\\
  \hline
  Data                             & \multicolumn{1}{c}{183}               & \multicolumn{1}{c}{139}     \\
  \end{scotch}
  \label{tab:evt}
\end{table}

Since there is no sign of an excess above the total background predictions,
it is possible to set only an upper limit on $\PW\PW\gamma$ and $\PW\Z\gamma$
cross sections, given the size of the current event sample.
The limit is calculated from the
event yields in Table~\ref{tab:evt} using a profile likelihood
asymptotic approximation method~(Appendix A.1.3 in Ref.~\cite{CMS-NOTE-2011-005}, \cite{Cowan:2010st}).
 An observed upper limit of 311\unit{fb} is calculated for the inclusive
cross section at 95\% confidence level (\CL), which is about 3.4 times larger than the standard model prediction of $91.6 \pm 21.7$\unit{fb} (with photon
$\ET > 30\GeV$ and $\abs{\eta} < 1.44$), calculated with {\sc aMC@NLO}. The
expected limit is 403\unit{fb} (4.4 times the SM).

\section{Limits on anomalous quartic couplings} 
\label{sec:limits_pT}

\begin{figure}[htb]
   \centering
\includegraphics[width=.48\textwidth]{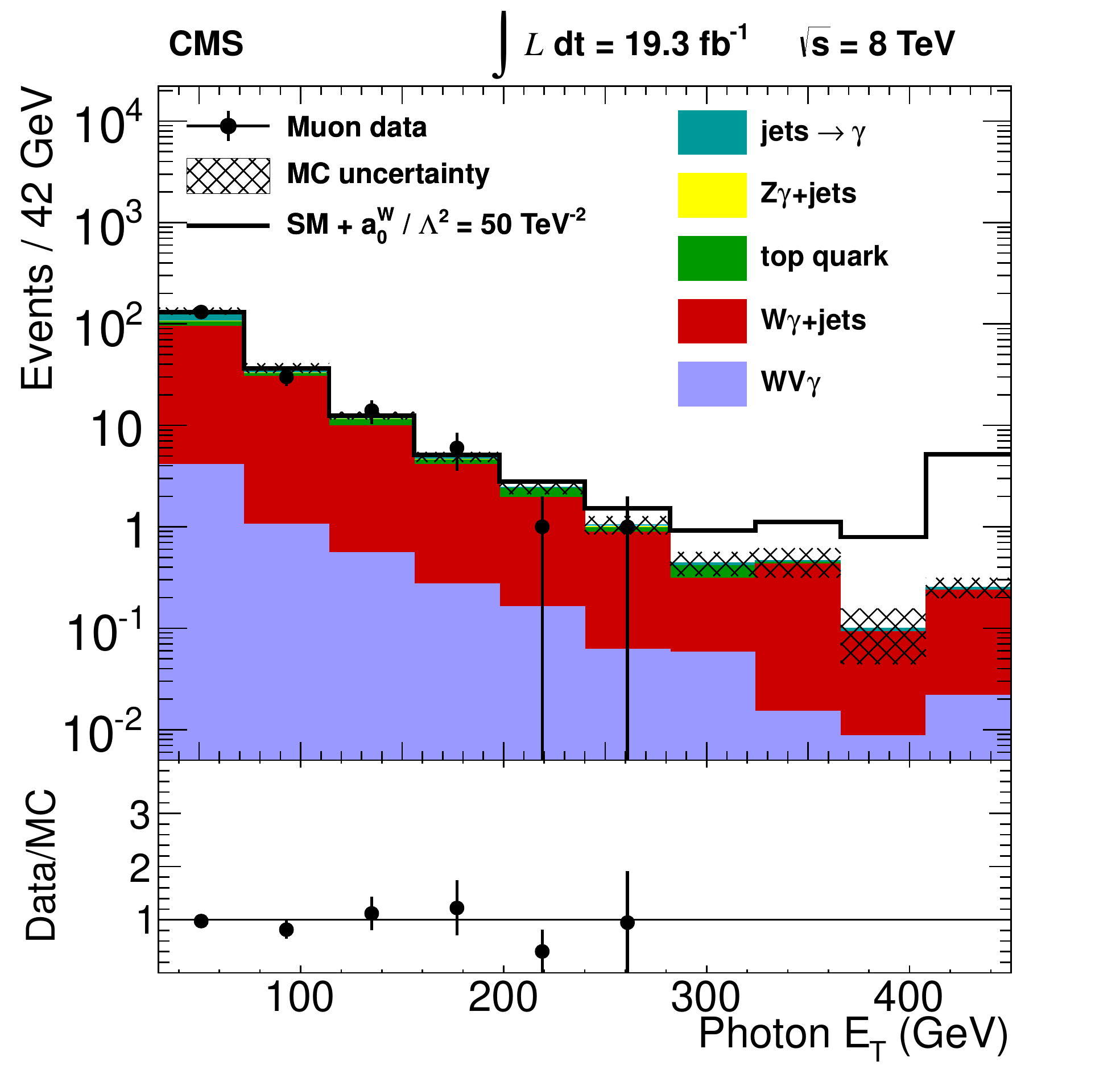}
\includegraphics[width=.48\textwidth]{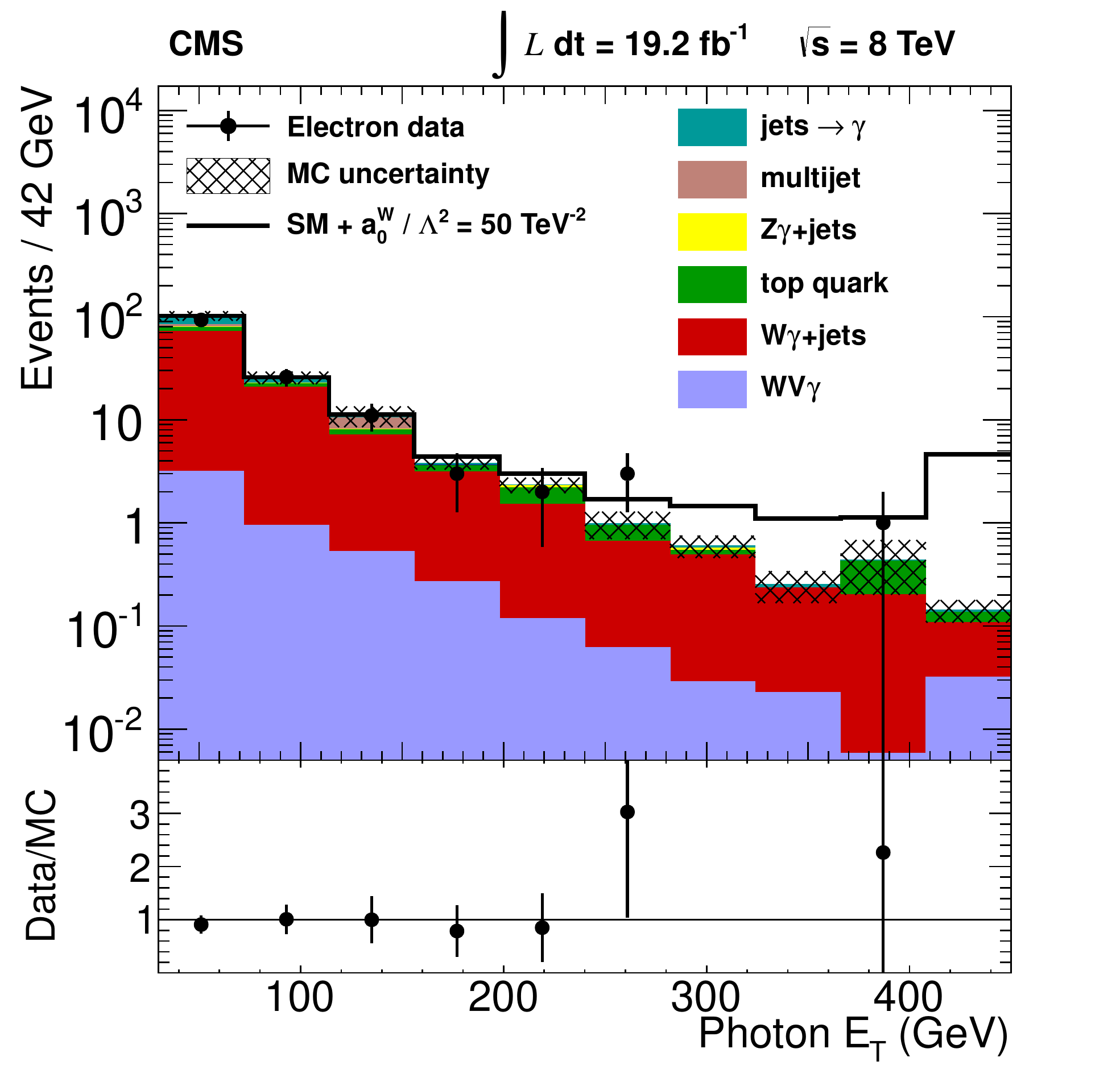}
\caption{Comparison of predicted and observed photon \ET
distributions in the (\cmsLeft) muon and (\cmsRight) electron channels. The rightmost bin
includes the integral of events above 450\GeV for each process. The solid black
line depicts a representative signal distribution with anomalous coupling parameter
$a_{0}^\PW/\Lambda^{2} = 50\TeV^{-2}$. }
\label{photonplots}
\end{figure}

The photon \ET distribution is sensitive to AQGCs and is therefore used to set limits on
the anomalous coupling parameters. Following the application
of all selection criteria, the photon \ET distributions for data, the total background, and the individual signal models
for the muon and electron channels are binned over the range 30--450\GeV. The photon \ET distributions for muon and electron
channels are shown in Fig.~\ref{photonplots}, along with the predicted signal from
$\PW\PW\gamma\gamma$ AQGC for $a_{0}^\PW/\Lambda^{2} = 50\TeV^{-2}$. The last bin includes the overflow.

The upper limits are set utilizing a profile likelihood asymptotic
approximation method~(Appendix A.1.3 in
Ref.~\cite{CMS-NOTE-2011-005}, \cite{Cowan:2010st}), which takes the
distributions from the two channels as independent inputs to be
combined statistically into a single result. Each coupling parameter
is varied over a set of discrete values, keeping the other parameters fixed
to zero; this causes the signal distribution to be altered
accordingly. The expected and observed signal strengths
$\sigma_{\text{excluded}}/\sigma_{\mathrm{AQGC}}$ are then calculated and
plotted against the corresponding coupling parameter values.

\begin{figure*}[!htb]
  \centering
    \includegraphics[width=0.49\textwidth]{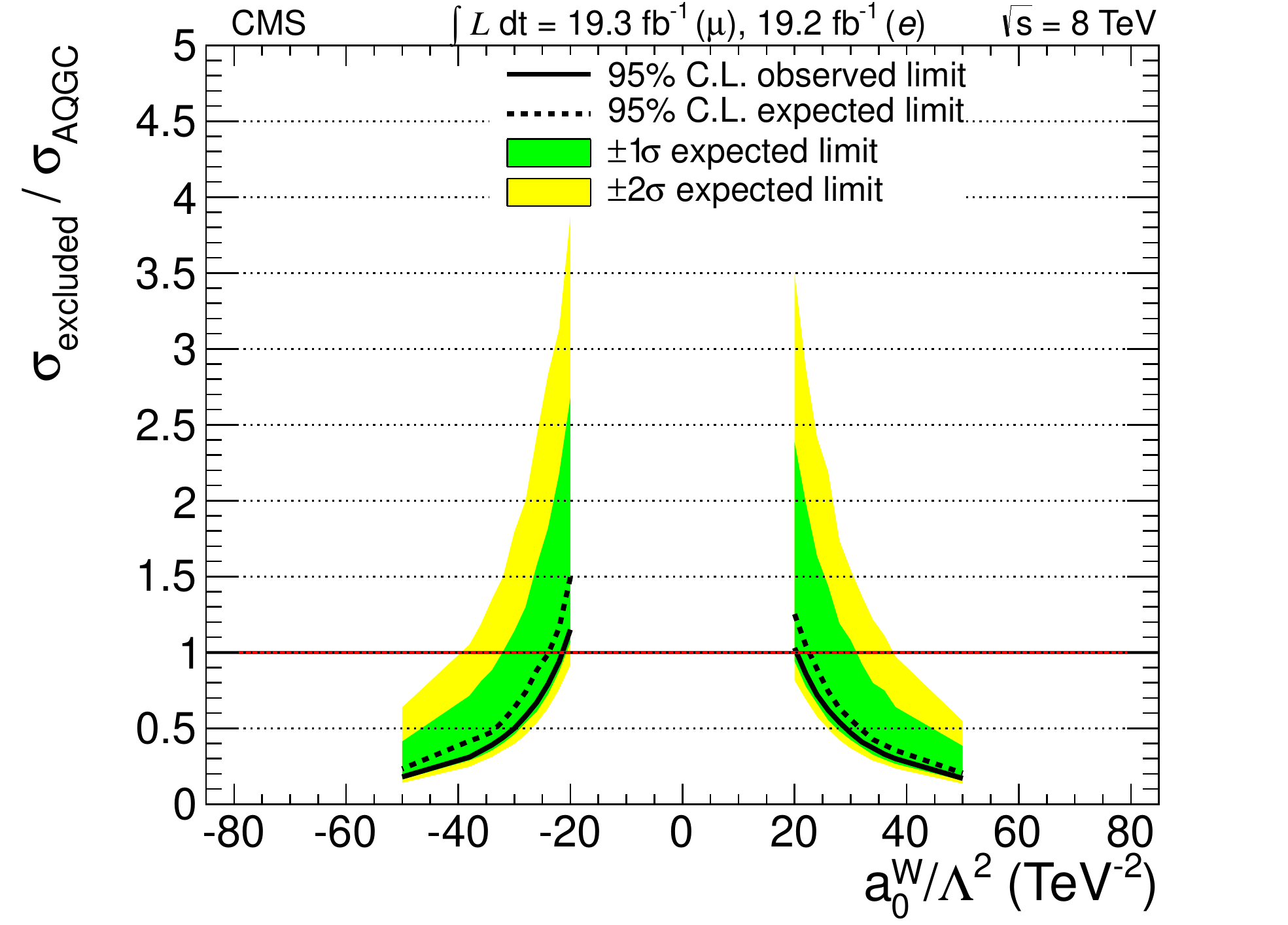}
    \includegraphics[width=0.49\textwidth]{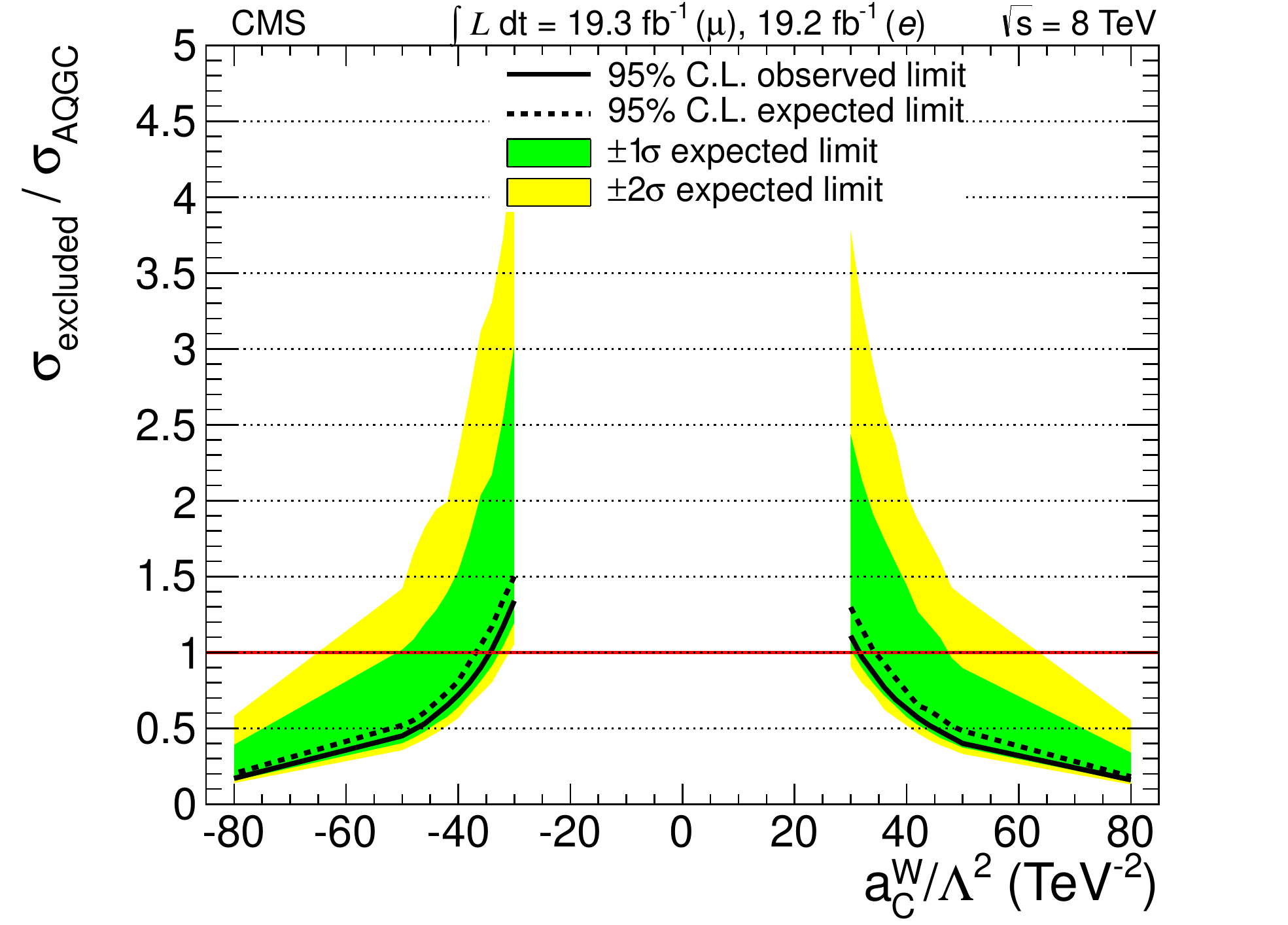}
    \includegraphics[width=0.49\textwidth]{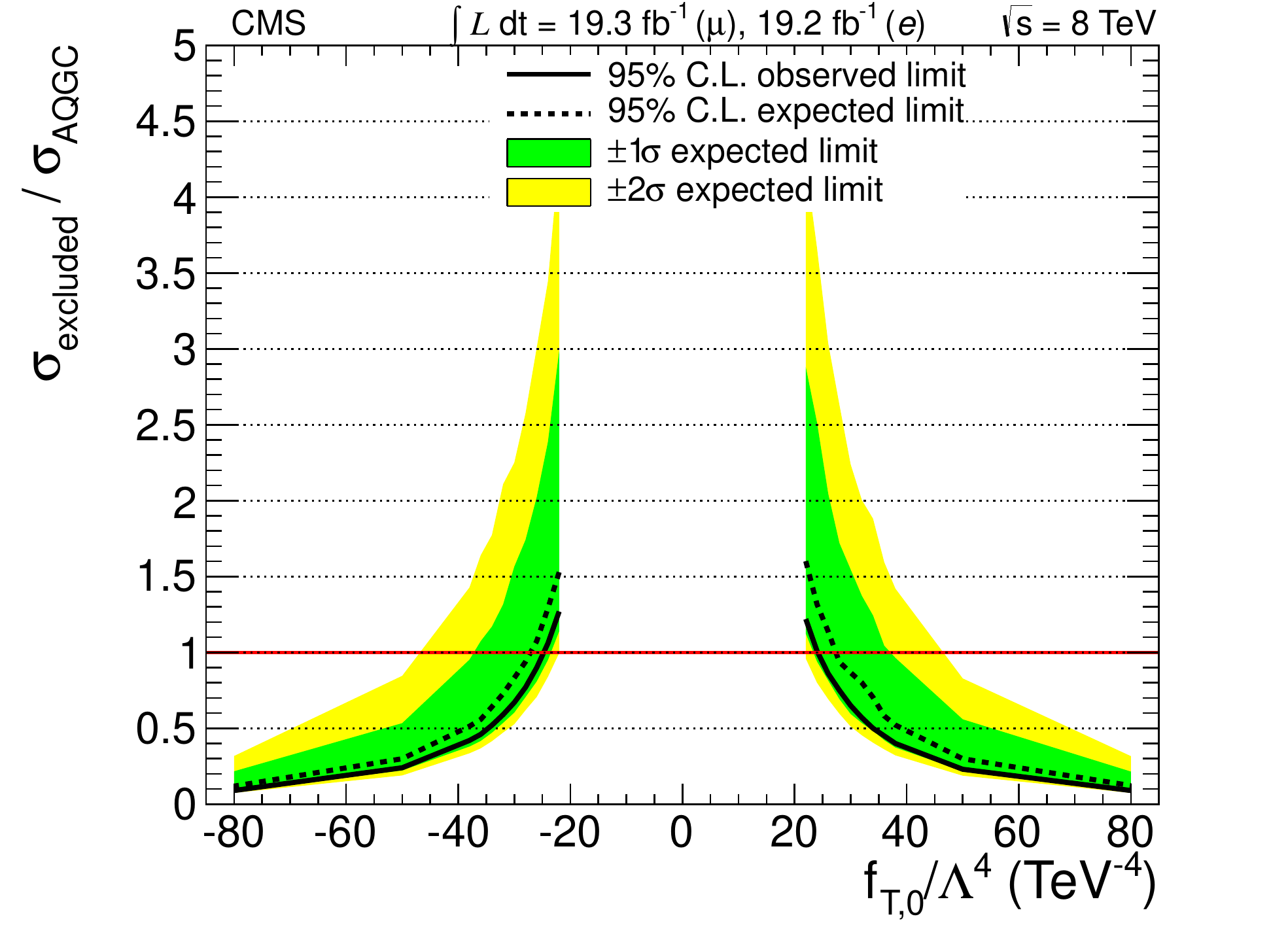}
    \includegraphics[width=0.49\textwidth]{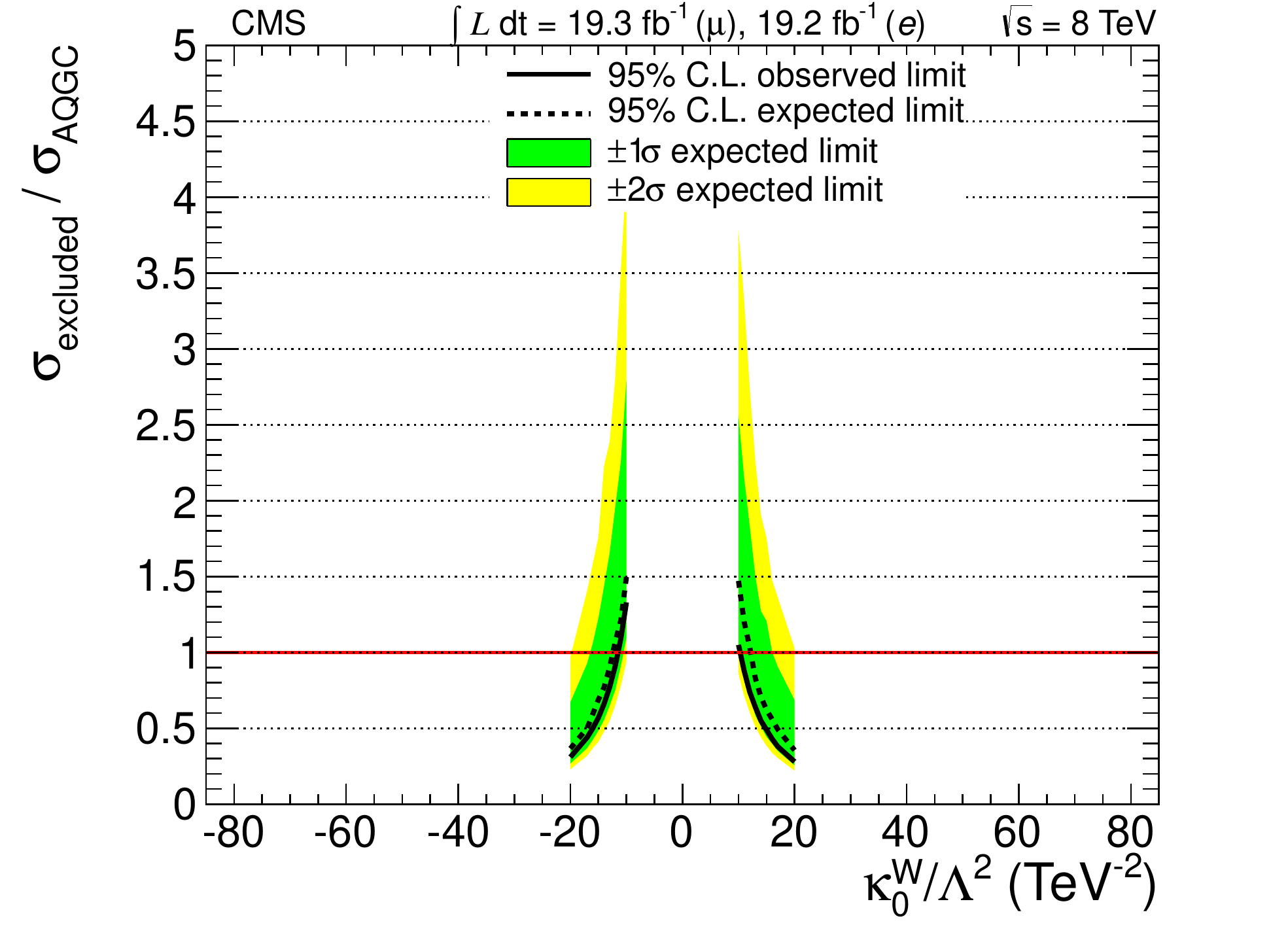}
    \includegraphics[width=0.49\textwidth]{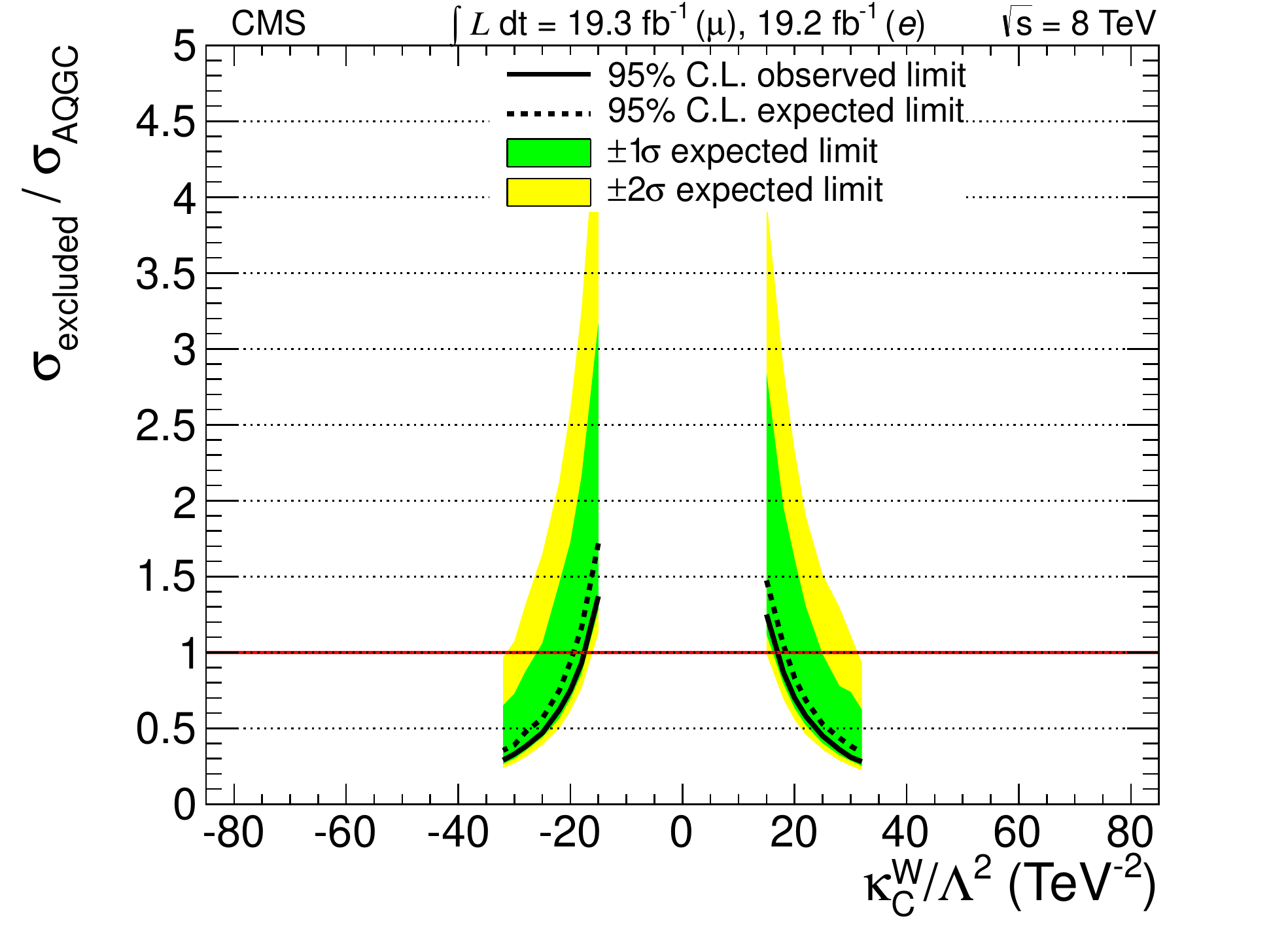}\\
    \caption{ 95\% \CL exclusion limits for (upper left) $a_{0}^\PW/\Lambda^{2}$, (upper right) $a_{C}^\PW/\Lambda^{2}$, (middle left) $f_{T,0}/\Lambda^{4}$, (middle right)
$\kappa_{0}^\PW/\Lambda^{2}$, and (bottom) $\kappa_{C}^\PW/\Lambda^{2}$.}
    \label{fig:limitshape1d_noMVA}
\end{figure*}

Figure~\ref{fig:limitshape1d_noMVA} shows the observed and expected exclusion limits for the
combination of muon and electron channels. Some positive/negative asymmetry is noticeable
in the plots because of SM/AQGC interference terms in the Lagrangian. Exclusion limits for
$a_{0}^\PW/\Lambda^{2}$, $a_{C}^\PW/\Lambda^{2}$,
$f_{T,0}/\Lambda^{4}$, $\kappa_{0}^\PW/\Lambda^{2}$, and
$\kappa_{C}^\PW/\Lambda^{2}$ are computed at 95\% \CL,
and are listed in Table~\ref{tab:limit_values_noMVA}.
Table~\ref{tab:limit_values_noMVA_dim8} reports the transformed
dimension-8 limits from the limits on the $a_{0}^\PW$ and $a_{C}^\PW$ parameters.

\begin{table}[htb]
  \topcaption{The 95\% \CL exclusion limits for each AQGC parameter from the combination of the muon and electron channels.}
  \centering
  \begin{scotch}{ll}
  \multicolumn{1}{c}{Observed limits} & \multicolumn{1}{c}{Expected limits} \\
  \hline
  $-21 < a_{0}^\PW/\Lambda^{2} < 20\TeV^{-2}$      & $-24  < a_{0}^\PW/\Lambda^{2} < 23\TeV^{-2}$  \\
  $-34 < a_{C}^\PW/\Lambda^{2} < 32\TeV^{-2}$      & $-37  < a_{C}^\PW/\Lambda^{2} < 34\TeV^{-2}$  \\
  $-25 < f_{T,0}/\Lambda^{4} < 24\TeV^{-4}$         & $-27  < f_{T,0}/\Lambda^{4} < 27\TeV^{-4}$ \\
  $-12 < \kappa_{0}^\PW/\Lambda^{2} < 10\TeV^{-2}$  & $-12  < \kappa_{0}^\PW/\Lambda^{2} < 12\TeV^{-2}$  \\
  $-18 < \kappa_{C}^\PW/\Lambda^{2} < 17\TeV^{-2}$  & $-19  < \kappa_{C}^\PW/\Lambda^{2} < 18\TeV^{-2}$  \\
  \end{scotch}
  \label{tab:limit_values_noMVA}
\end{table}

\begin{table}[htb]
 \centering
  \topcaption{The 95\% \CL exclusion limits for each dimension-8 AQGC parameter from the combination of the muon and electron channels.}
  \begin{scotch}{ll}
  \multicolumn{1}{c}{Observed limits ($\TeV^{-4}$)} & \multicolumn{1}{c}{Expected limits ($\TeV^{-4}$)} \\
  \hline
  $-77  < f_{M,0}/\Lambda^{4} < 81$  & $-89  < f_{M,0}/\Lambda^{4} < 93$  \\
  $-131 < f_{M,1}/\Lambda^{4} < 123$ & $-143 < f_{M,1}/\Lambda^{4} < 131$ \\
  $-39  < f_{M,2}/\Lambda^{4} < 40$  & $-44  < f_{M,2}/\Lambda^{4} < 46$  \\
  $-66  < f_{M,3}/\Lambda^{4} < 62$  & $-71  < f_{M,3}/\Lambda^{4} < 66$  \\
  \end{scotch}
  \label{tab:limit_values_noMVA_dim8} \end{table}

\begin{figure}[!htb]
  \centering
      \includegraphics[width=\cmsFigWidth]{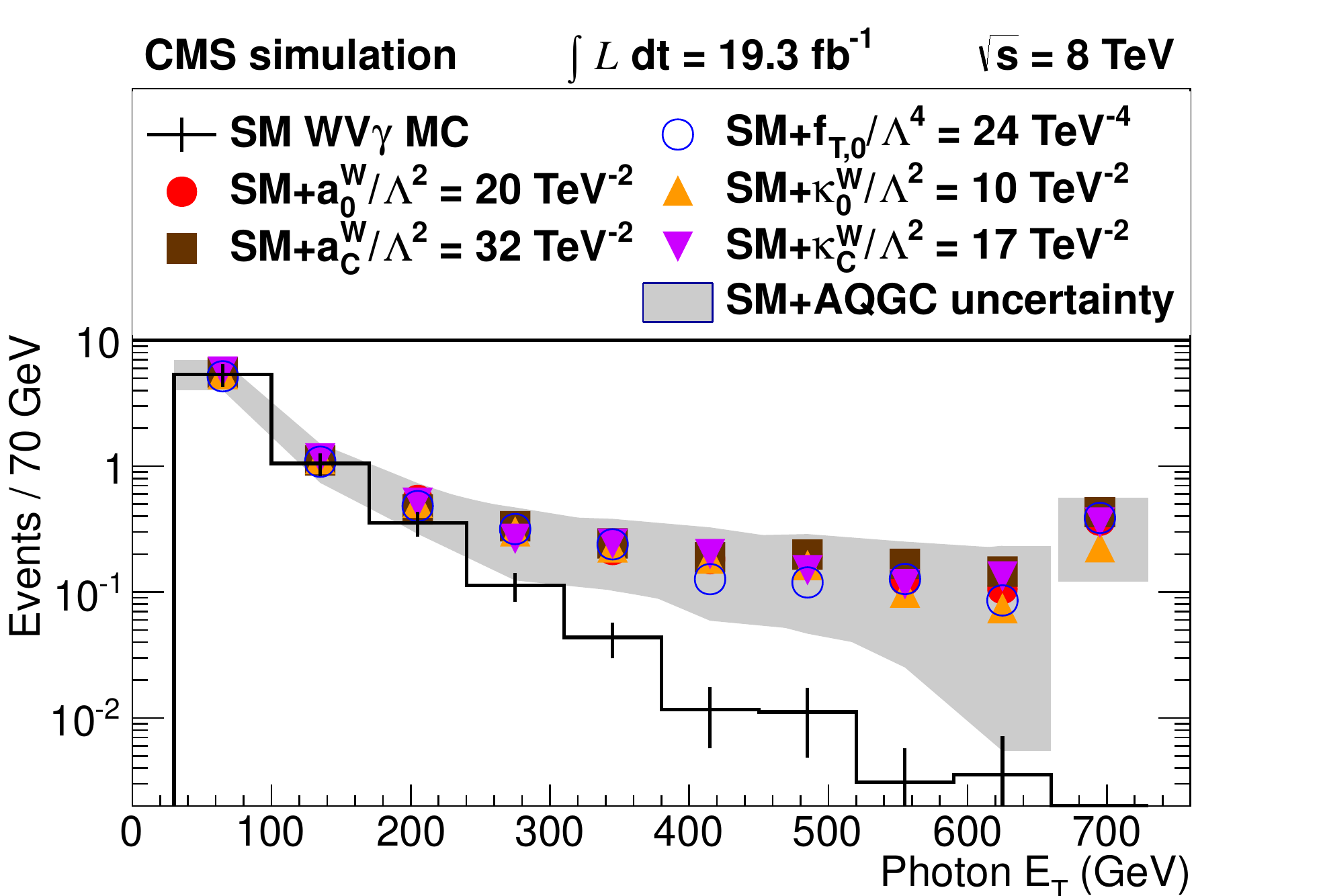}
   \caption{Expected photon \ET distributions after the selection for the muon
channel is applied: SM prediction, SM plus AQGC prediction for
$a_{0}^\PW/\Lambda^{2}$, $a_{C}^\PW/\Lambda^{2}$,
$f_{T,0}/\Lambda^{4}$, $\kappa_{0}^\PW/\Lambda^{2}$, and
$\kappa_{C}^\PW/\Lambda^{2}$. Systematic and statistic uncertainties are shown. 
The last bin includes the overflow. }
    \label{fig:limitinput}
\end{figure}

Figure~\ref{fig:limitinput} shows the photon \ET distributions for a signal in the muon channel corresponding to AQGC
parameters that are set to the limits we have obtained. The distributions for the various AQGC values are similar. The contribution from AQGC is 
prominent in the region $\ET > 240\GeV$, where the expected number of signal events is approximately 1.4. 
The corresponding distributions for the electron channel are similar.

\begin{figure*}[bhtp]
\centering
   \includegraphics[width=0.9\textwidth]{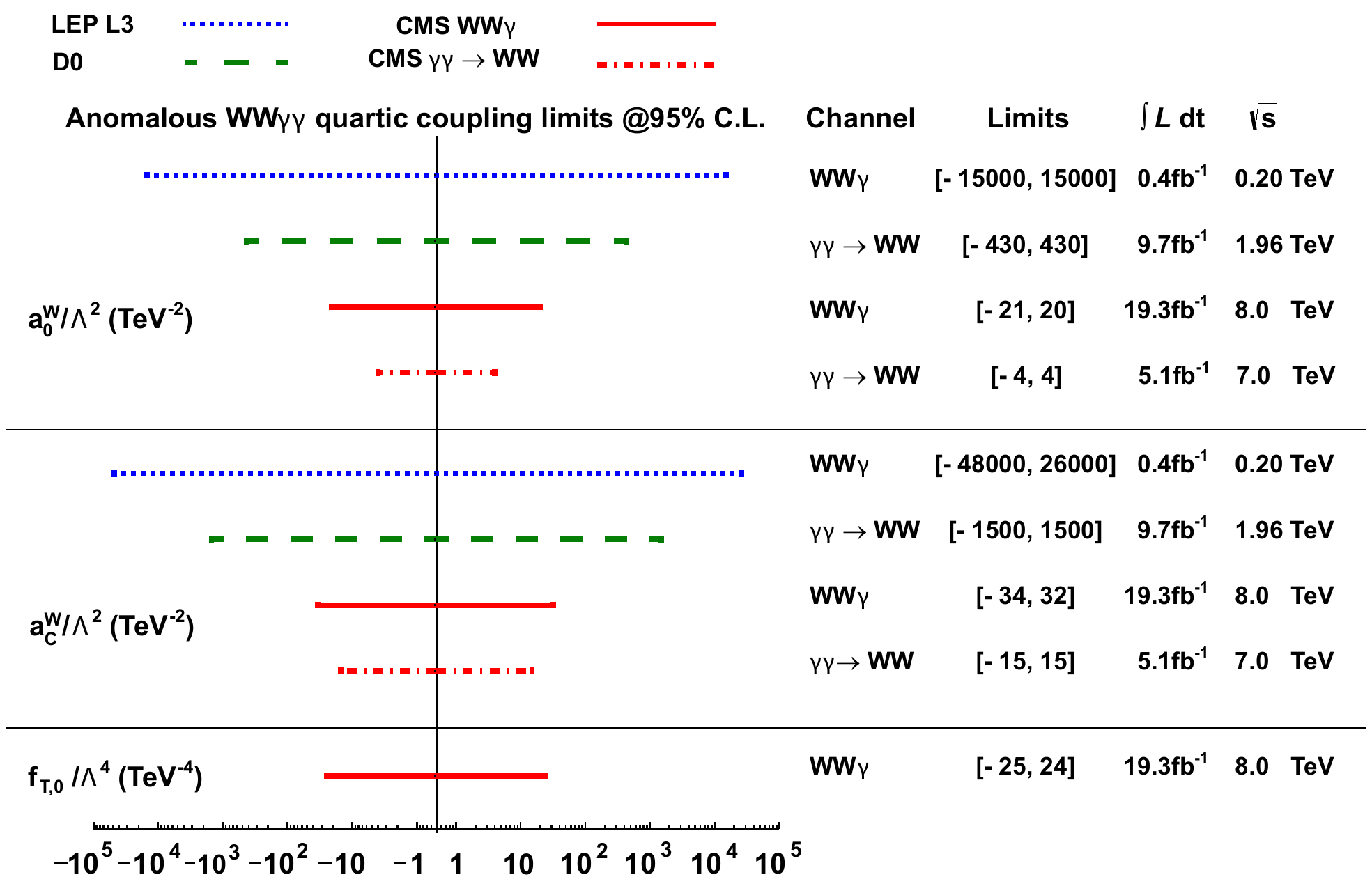}
   \caption{ Comparison of the limits on the $\PW\PW\gamma\gamma$ AQGC parameter obtained
from this study, together with results from exclusive $\gamma\gamma\to${\PW\PW} production at CMS~\cite{Chatrchyan:2013foa} and results from the
L3~\cite{Achard:2001eg} and the D0~\cite{Abazov:2013opa} collaborations. All limits on AQGC are calculated without a form
factor. }
\label{fig:WWAAcomparison}
\end{figure*}

A comparison of several existing limits on the $\PW\PW\gamma\gamma$ AQGC parameter is
shown in Fig.~\ref{fig:WWAAcomparison}. Existing limits include the result from
exclusive $\gamma\gamma\to${\PW\PW} production at CMS~\cite{Chatrchyan:2013foa},
in addition to results from the L3~\cite{Achard:2001eg} and the D0~\cite{Abazov:2013opa}
collaborations. All of the limits shown on AQGC are calculated without a form factor.

\section{Summary}
\label{sec:summary}

   A search for ${\PW}V\gamma$ triple vector boson production that results in
   constraints on anomalous quartic gauge boson couplings has been presented
   using events containing a {\PW} boson decaying to leptons, a second
   boson $V$ ($V$ = {\PW} or {\Z}) boson, and a photon. The data
   analyzed correspond to an integrated luminosity of 19.3\fbinv
   collected in $\Pp\Pp$ collisions at $\sqrt{s} = 8\TeV$ in 2012 with the
   CMS detector at the LHC. An upper limit of 311\unit{fb} at 95\% \CL is obtained for
   the production of ${\PW}V\gamma$ with photon $\ET > 30\GeV$ and $\abs{\eta} < 1.44$.
   No evidence for anomalous $\PW\PW\gamma\gamma$ and $\PW\PW\Z\gamma$ quartic
   gauge couplings is found. The following constraints are obtained for these
   couplings at 95\% \CL:

\begin{align*}
   -21 &< a_{0}^\PW/\Lambda^{2} < 20\TeV^{-2}, \\
   -34 &< a_{C}^\PW/\Lambda^{2} < 32\TeV^{-2}, \\
   -25 &< f_{T,0}/\Lambda^{4} < 24\TeV^{-4}, \\
   -12 &< \kappa_{0}^\PW/\Lambda^{2} < 10\TeV^{-2},\quad  \text{and} \\
   -18 &< \kappa_{C}^\PW/\Lambda^{2} < 17\TeV^{-2}.
\end{align*}

   These are the first experimental limits reported on $f_{T,0}$ and the
   CP-conserving couplings $\kappa_0^\PW$ and $\kappa_C^\PW$.
   Figure~\ref{fig:WWAAcomparison} compares the constraints on the
   $\PW\PW\gamma\gamma$ AQGC parameter obtained from this study with those
   obtained in previous analyses.
\section*{Acknowledgments}

We congratulate our colleagues in the CERN accelerator departments for the excellent performance of the LHC and thank the technical and administrative staffs at CERN and at other CMS institutes for their contributions to the success of the CMS effort. In addition, we gratefully acknowledge the computing centres and personnel of the Worldwide LHC Computing Grid for delivering so effectively the computing infrastructure essential to our analyses. Finally, we acknowledge the enduring support for the construction and operation of the LHC and the CMS detector provided by the following funding agencies: BMWFW and FWF (Austria); FNRS and FWO (Belgium); CNPq, CAPES, FAPERJ, and FAPESP (Brazil); MES (Bulgaria); CERN; CAS, MoST, and NSFC (China); COLCIENCIAS (Colombia); MSES and CSF (Croatia); RPF (Cyprus); MoER, SF0690030s09 and ERDF (Estonia); Academy of Finland, MEC, and HIP (Finland); CEA and CNRS/IN2P3 (France); BMBF, DFG, and HGF (Germany); GSRT (Greece); OTKA and NIH (Hungary); DAE and DST (India); IPM (Iran); SFI (Ireland); INFN (Italy); NRF and WCU (Republic of Korea); LAS (Lithuania); MOE and UM (Malaysia); CINVESTAV, CONACYT, SEP, and UASLP-FAI (Mexico); MBIE (New Zealand); PAEC (Pakistan); MSHE and NSC (Poland); FCT (Portugal); JINR (Dubna); MON, RosAtom, RAS and RFBR (Russia); MESTD (Serbia); SEIDI and CPAN (Spain); Swiss Funding Agencies (Switzerland); MST (Taipei); ThEPCenter, IPST, STAR and NSTDA (Thailand); TUBITAK and TAEK (Turkey); NASU and SFFR (Ukraine); STFC (United Kingdom); DOE and NSF (USA).

Individuals have received support from the Marie-Curie programme and the European Research Council and EPLANET (European Union); the Leventis Foundation; the A. P. Sloan Foundation; the Alexander von Humboldt Foundation; the Belgian Federal Science Policy Office; the Fonds pour la Formation \`a la Recherche dans l'Industrie et dans l'Agriculture (FRIA-Belgium); the Agentschap voor Innovatie door Wetenschap en Technologie (IWT-Belgium); the Ministry of Education, Youth and Sports (MEYS) of Czech Republic; the Council of Science and Industrial Research, India; the Compagnia di San Paolo (Torino); the HOMING PLUS programme of Foundation for Polish Science, cofinanced by EU, Regional Development Fund; and the Thalis and Aristeia programmes cofinanced by EU-ESF and the Greek NSRF.

\bibliography{auto_generated}   

\cleardoublepage \appendix\section{The CMS Collaboration \label{app:collab}}\begin{sloppypar}\hyphenpenalty=5000\widowpenalty=500\clubpenalty=5000\input{SMP-13-009-authorlist.tex}\end{sloppypar}
\end{document}

%% file: SMP-13-009-authorlist.tex
\textbf{Yerevan Physics Institute,  Yerevan,  Armenia}\\*[0pt]
S.~Chatrchyan, V.~Khachatryan, A.M.~Sirunyan, A.~Tumasyan
\vskip\cmsinstskip
\textbf{Institut f\"{u}r Hochenergiephysik der OeAW,  Wien,  Austria}\\*[0pt]
W.~Adam, T.~Bergauer, M.~Dragicevic, J.~Er\"{o}, C.~Fabjan\cmsAuthorMark{1}, M.~Friedl, R.~Fr\"{u}hwirth\cmsAuthorMark{1}, V.M.~Ghete, C.~Hartl, N.~H\"{o}rmann, J.~Hrubec, M.~Jeitler\cmsAuthorMark{1}, W.~Kiesenhofer, V.~Kn\"{u}nz, M.~Krammer\cmsAuthorMark{1}, I.~Kr\"{a}tschmer, D.~Liko, I.~Mikulec, D.~Rabady\cmsAuthorMark{2}, B.~Rahbaran, H.~Rohringer, R.~Sch\"{o}fbeck, J.~Strauss, A.~Taurok, W.~Treberer-Treberspurg, W.~Waltenberger, C.-E.~Wulz\cmsAuthorMark{1}
% ...
\vskip\cmsinstskip
% ...
1:~~Also at Vienna University of Technology, Vienna, Austria\\
2:~~Also at CERN, European Organization for Nuclear Research, Geneva, Switzerland\\
% ...
